\documentclass[prb,aps,reprint,showpacs,twocolumn]{revtex4-1}
\usepackage{graphicx,amsmath}
\usepackage{float}
\usepackage{subfigure}
\usepackage{mathrsfs}
\usepackage{graphics}
\usepackage{epstopdf}
\usepackage{amsmath}
\usepackage{amssymb}
\usepackage{hyperref}
\usepackage{color}
\usepackage{braket}
\begin{document}

\author{Kallol Mondal and Charudatt Kadolkar}
\affiliation{Department of Physics, Indian Institute of Technology Guwahati, Guwahati, Assam 781039, India}
\date{\today}
\title{Schwinger Boson mean field theory of kagome Heisenberg antiferromagnet with Dzyaloshinskii-Moriya interaction}
\begin{abstract}
We have studied the effect of the  Dzyaloshinskii-Moriya interaction on kagome Heisenberg antiferromagnet  using  Schwinger boson mean field theory(SBMFT).  Within SBMFT framework, Messio et al had argued that the ground state of kagome antiferromagnet is possibly a chiral topological spin liquid(Phys. Rev. Lett. 108, 207204 (2012)). Thus, we have computed zero-temperature ground state phase diagram considering the time-reversal breaking states as well as fully symmetric \textit{Ans\"{a}tze}. We discuss the relevance of these results in experiments and other studies. Finally, we have computed the static and dynamic spin structure factors in relevant phases.
\end{abstract}
\maketitle
\section{\label{sec:intro}Introduction}
Quantum spin liquids (QSL)~\cite{anderson1973resonating} are the exotic states of matter without any broken symmetries even at $T=0$, the  states of matter which cannot be explained in the paradigm of Landaus's symmetry breaking theory~\cite{lhuillier2011introduction,Mila2000,balents2010spin,balents2010spin,savary2016quantum}. The most promising candidate to posses spin liquid ground states is spin-1/2 kagome Heisenberg antiferromagnet (KHAF) due to its low-dimension, small-spin, and strong geometrical frustration~\cite{PhysRevB.47.5459,yan2011spin,PhysRevLett.101.117203,PhysRevLett.98.117205}. One of the challenges of the search for QSL among materials is the presence of anisotropies like Dzyaloshinskii-Moriya (DM) interaction. Such interactions reduce symmetry and quantum fluctuations, and lead to magnetic ordering. The Dzyaloshinskii-Moriya interaction ~\cite{dzyaloshinskii1957thermodynamic,moriya1960anisotropic}, arises when there is lack inversion symmetry in the lattice.

The mineral, Herbertsmithite, is an example where the spin-1/2 copper atoms form a kagome lattice. No magnetic order has been found down  to 50 mK with the exchange coupling being 170K~\cite{helton2007spin,lee2008end}. To explain the spin susceptibility enhancement of Herbertsmithite~\cite{levi2007new}at low temperature,  a small value of Dzyaloshinskii-Moriya interaction has to be considered. Presence of DM interaction is also confirmed by paramagnetic resonance~\cite{zorko2008dzyaloshinsky} where the DM interaction of strength of $0.08J$ is needed to explain the line width. Small values of DM interaction can produce long range order in a spin system, specially in spin liquid ground states. So it was not clear why spins do not freeze in Herbertsmithite at low temperatures.

The effect of DM interaction on the kagome Heisenberg antiferromagnet has been studied by several groups~\cite{rigol2007magnetic, rousochatzakis2009dzyaloshinskii,cepas2008quantum,tovar2009dzyaloshinskii,hermele2008properties,messio2010schwinger}. From exact diagonalization(ED) Cepas et. al~\cite{cepas2008quantum}, they found that there is a critical point at $D_{c}=0.1J$, where there is moment free phase at the lower side and Neel phase on the other. This result is consistent with experiments since estimated DM interaction strength is about $0.08J$ in Herbertsmithite. Sachdev has proposed a quantum critical theory~\cite{huh2010quantum} about the critical point suggested by Cepas. Messio et al,~\cite{messio2010schwinger} using Schwinger-boson mean field theory (SBMFT), have calculated the phase diagram and showed that the results are qualitatively similar to the ED studies in small boson density region.

The purpose of this paper is two-fold. In 2013, Messio et al.~\cite{messio2012kagome} showed that in SBMFT framework, that the ground state of the Kagome Heisenberg antiferromagnet is chiral $\mathbb{Z}_{2}$ spin liquid. In their study of DM interaction, they had not considered the time-reversal breaking \textit{Ans\"{a}tze}~\cite{messio2010schwinger}. This would change the phase diagram for small strength of DM interaction. Secondly, they have considered only bond creation mean field $\mathcal{A}$. However, the inclusion of a second boson hopping mean field, $\mathcal{B}$ field improves the quantitative agreement of the band width in the excitation spectrum of the Heisenberg model on triangular lattice~\cite{mezio2011test}. Flint and Coleman showed that these two fields together give the
better description of these frustrated systems~\cite{flint2009symplectic}. Thus, we, in this paper have considered chiral spin liquid \textit{Ansatz} and also included the hopping mean field.

In Sec.II we present the model and the brief of Schwinger boson mean field theory  with two bond operators.  In Sec.III we have described the detailed algorithm for the numerical search of the optimum points.  In the Sec.IV we present the zero temperature ground state phase diagram with properties associated with it and effect of DM interaction on the QSL ground state. We have discussed the connection between the prediction of our model with experimental results.  We have also calculated the dynamical structure factor of the different  \textit{Ans\"{a}tze} in relevant phases in the Sec.V.
\section{Model And Formalism}
\begin{figure}
	\includegraphics[height=4cm,width=7cm]{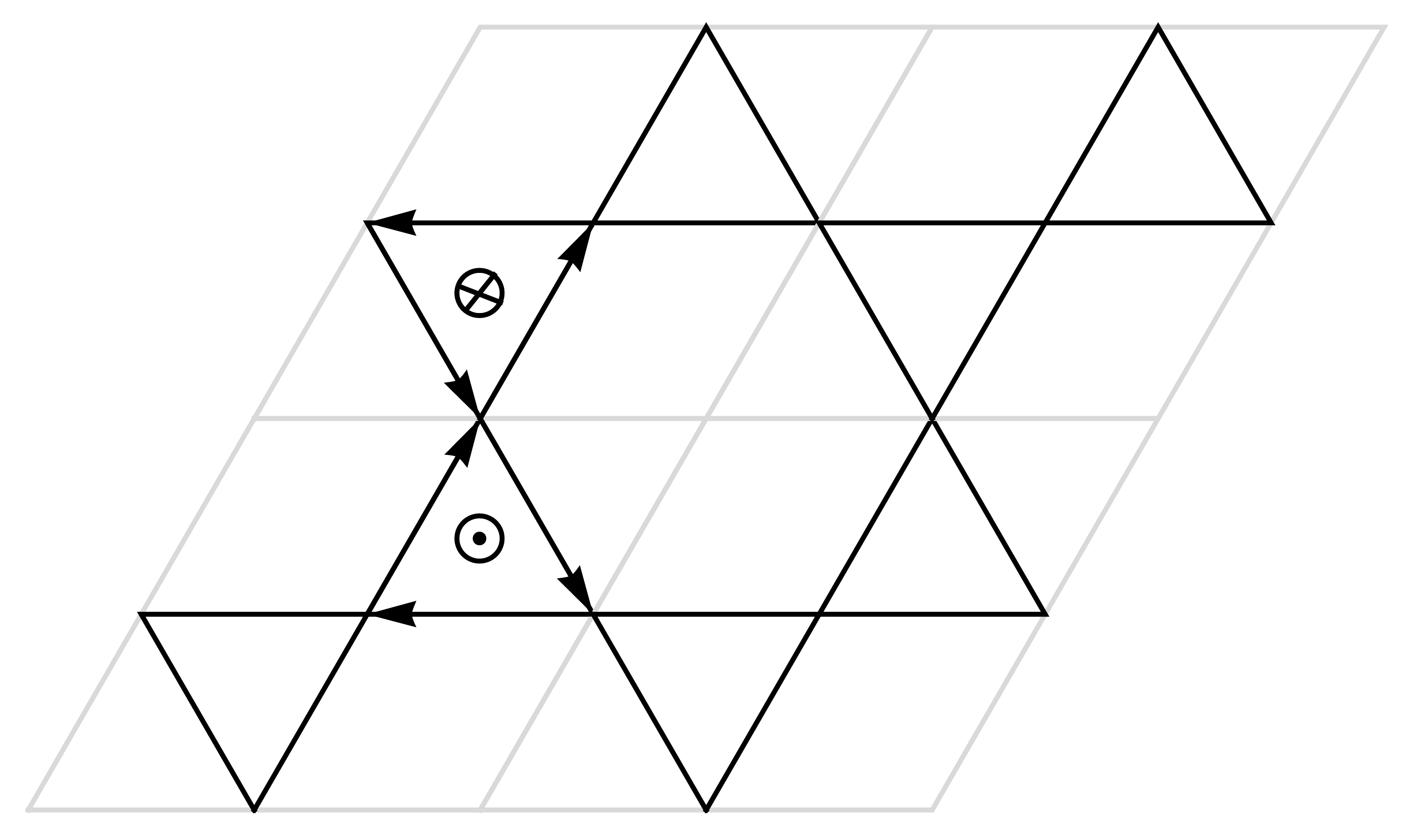}
	\caption{Bond directions for Dzyaloshinskii-Moriya interaction  where the DM vector $\vec{D}$ (shown at the center of triangle) is staggered between up to down triangles.}
	\label{fig:dmdiagram}
\end{figure}

The Hamiltonian for  nearest neighbour Heisenberg model with  Dzyaloshinskii-Moriya interaction is given by
\begin{equation}
 	H = \sum_{\langle ij \rangle}\big[J \vec{S_i}\cdot{\vec{S_j}} +  \vec{D}_{ij}\cdot(\vec{S_i}\times{\vec{S_j}})\big]
\end{equation}
with  $ \langle ij \rangle$ stands for a pair of neighbouring sites $i$ and $j$. The pairs are directed due to the DM interaction and the directions are shown in Fig.~\ref{fig:dmdiagram}. The Heisenberg coupling is assumed to be antiferromagnetic ( $ J > 0 $ ). The DM vector is taken to be perpendicular to the plane of the lattice as shown in the figure with constant magnitude $D$. The planar component of DM vector can be taken care of by rotating the spins appropriately as long as the planar component is small~\cite{cepas2008quantum}. Introduction of the DM interaction reduces the global symmetry of the Heisenberg hamiltonian from $SU(2)$ to $U(1)$. However, the wallpaper group remains $P6m$. 
\begin{figure*}
	\includegraphics[height=4cm,width=8.5cm]{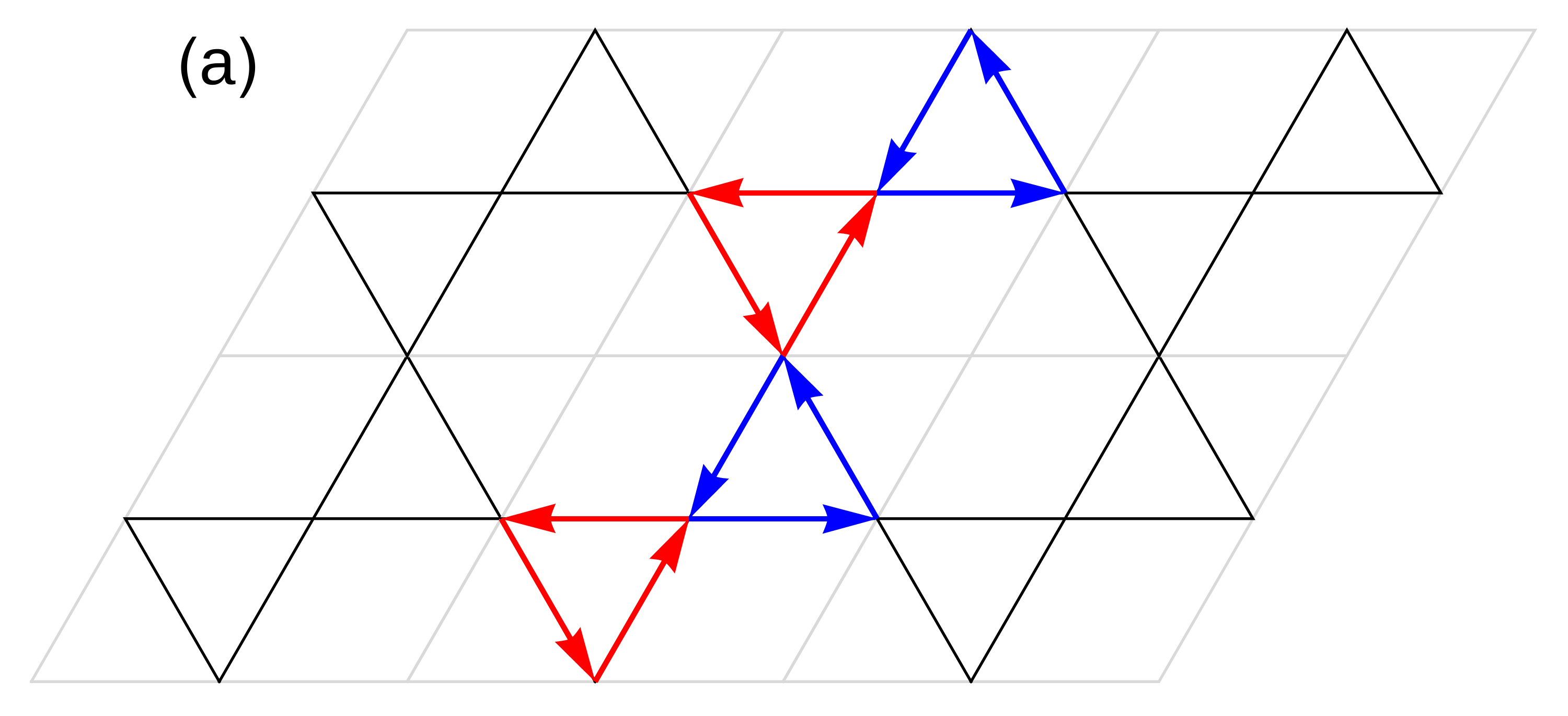}
	\includegraphics[height=4cm,width=8.5cm]{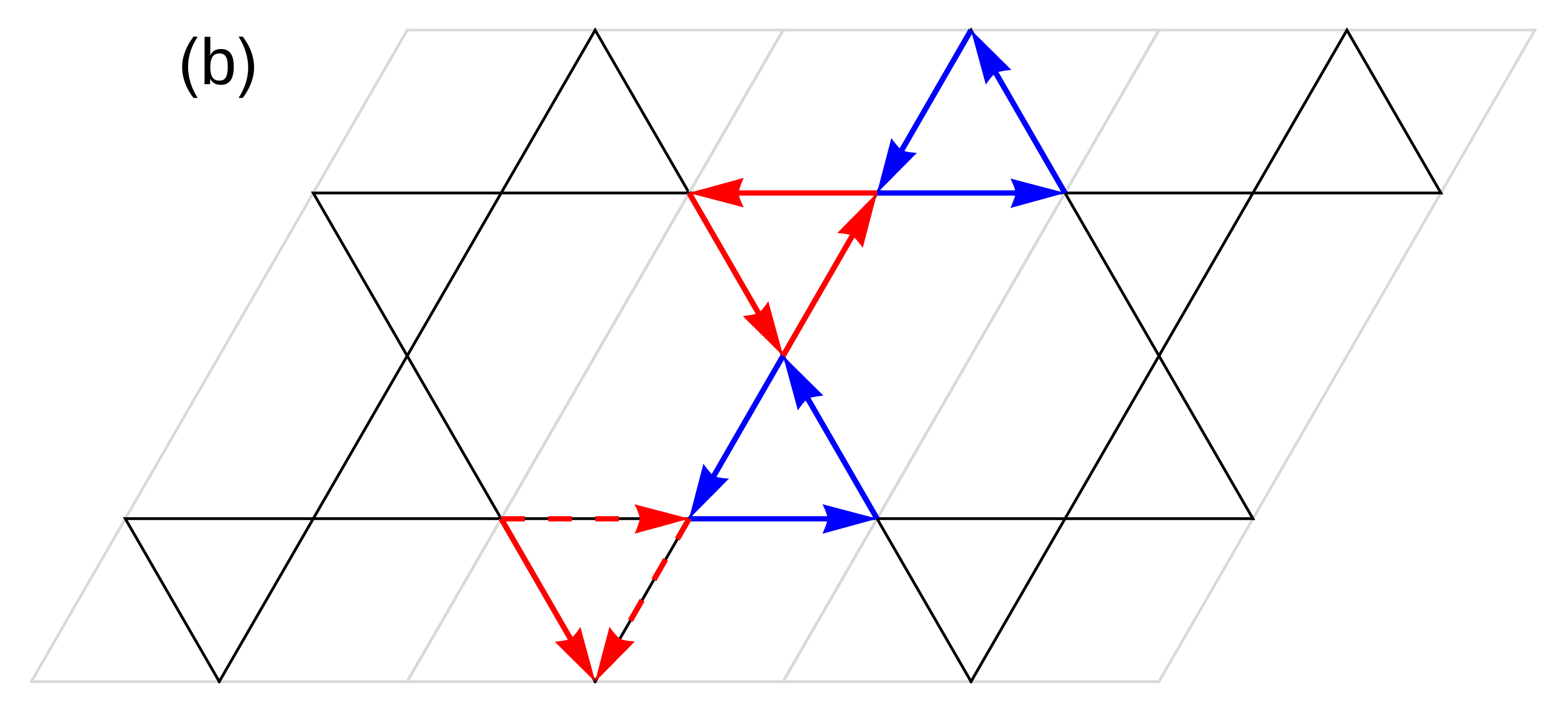}
	\caption{Grey lines mark the three-site unit cell for the $(0,0)$,$(\pi,0)$  
	\textit{Ansatz} in the Fig. (a) and six-site unit cell for \textit{cuboc1} \textit{Ansatz} in Fig. (b). The mean fields are defined as $\langle \mathcal{A}_{ij} \rangle = |\mathcal{A}| e^{i \phi_{\mathcal{A}}}$ and  $\langle \mathcal{B}_{ij} \rangle = |\mathcal{B}| e^{i \phi_{\mathcal{B}}}$. In the Fig. (a), for $(0,0)$ \textit{Ansatz} $\phi_{\mathcal{A}} = 0$ for both the red (down) triangles and blue (up) triangles. For $(\pi,0)$ \textit{Ansatz} $\phi_{\mathcal{A}} = 0$ for red triangles and $\phi_{\mathcal{A}} = \pi$ for blue triangles. For both the  \textit{Ans\"{a}tze} $\phi_{\mathcal{B}}=\pi$ for all bonds.  In the Fig. (b), for $cuboc1$ \textit{Ansatz}, $(\phi_A,\phi_B)=(\phi,\pi)$ for blue triangles where $\phi$ is arbitrary. For  red triangles, for undashed lines $(\phi_A,\phi_B)=(0,\pi)$ and for dashed lines $(\phi_A,\phi_B)=(\pi,0)$ .}
 	\label{fig:ansatze}
\end{figure*}
It is expected that the DM interaction will induce long range order and hence using SBMFT approach is appropriate since it elegantly treats both ordered and spin liquid phases. In Schwinger boson representation the spin operators are mapped to the bosonic operators as follows,
\begin{eqnarray}
	S^\alpha_i &= & \frac{1}{2} \begin{pmatrix}a_i^\dagger& b_i^\dagger \end{pmatrix} \sigma^\alpha \begin{pmatrix}a_i \\ b_i \end{pmatrix} 
\end{eqnarray}
where $i$ is the site index and $\alpha ={x,y,z }$.  $\sigma^\alpha $ are the Pauli matrices. This mapping is faithful only when the number of bosons is equal to $2S$, that is,  $(a_i^\dagger a_i + b_i^\dagger b_i  ) = 2S$. This constraint, in principle, must be satisfied at each site. However, it is interesting to study the properties of the bosonic system by implementing this constraint on average basis by treating $2S$ as a parameter. Let $\kappa = \braket{n_i}/2$ be the boson density per flavor. The spin wave function is then obtained by applying appropriate projection operator.

 We define  bond creation operators as  $\hat{A}_{ij}^\dagger=\frac{1}{2}(e^{i \theta_{ij}}a_i^\dagger b_j^\dagger - e^{-i \theta_{ij}}b_i^\dagger a_j^\dagger)$ and $\hat{B}^\dagger_{ij}=\frac{1}{2}(a_i a^\dagger_j +b_i b^\dagger_j )$ with $\theta_{ij}=D_{ij}/J$. The operator $A^\dagger_{ij}$  creates a mixture of a singlet and a triplet on the bond, while $B^\dagger_{ij}$ is represents coherent hopping of the bosons between the sites. These definitions are slightly different from those used by Manuel et al~\cite{Manuel1996}. We can rewrite the  Hamiltonian in terms of these bond operators with an approximation that $D/J$ is very small as,
\begin{equation}
	H=\sum_{\langle ij \rangle} J \big[ : \hat{B}_{ij}^\dagger \hat{B}_{ij} : - \hat{A}_{ij}^\dagger \hat{A}_{ij} \big]
\end{equation}
where $::$ means normal order. We decouple the quartic term by introducing mean fields $ {\cal{A}}_{ij} = \left\langle \hat{A}_{ij} \right\rangle$ and $ {\cal{B}}_{ij} = \left\langle \hat{B}_{ij} \right\rangle$. Then the mean field hamiltonian is given by 
\begin{equation}
	H_{\text{MF}}=\sum_{\langle ij \rangle} J \big[ \mathcal{B}_{ij}\hat{B}_{ij}^\dagger -  \mathcal{A}_{ij} \hat{A}_{ij}^\dagger \big]+\text{H.C.} 
	- \sum_i \lambda_i \hat{n}_i +\epsilon_0
\end{equation}
where $\epsilon_0 = \sum_{\langle ij \rangle} J (\mathcal{A}_{ij}|^2 - |\mathcal{B}_{ij}|^2) + 2 S \sum_i \lambda_i$. The Lagrange's multiplier $\lambda_i$  has been added to constrain the mean boson number at each site.  

The PSG analysis of  $ \mathbb{Z}_2 $ spin liquids on kagome lattice was carried out by Wang and Vishwanath and they have shown that there are only four symmetric \textit{Ans\"{a}tze} possible~\cite{wang2006spin}. They have  labeled  these \textit{Ans\"{a}tze} as $(0,0), (\pi,0), (0,\pi)$ and $(\pi,\pi)$, based on the fluxes through the hexagon and the rhombus. These \textit{Ans\"{a}tze} respect time reversal symmetry and lead to planar long range order.  However, Messio et al have shown that ground state of Heisenberg antiferromagnet on kagome lattice is possibly a chiral spin liquid which is termed as {\it cuboc1}~\cite{messio2012kagome}. A systematic enumeration of {\it weakly symmetric} spin liquids shows that there are 20 families of spin liquid \textit{Ans\"{a}tze} on kagome lattice~\cite{messio2013time}. We restrict our choices to the four symmetric spin liquids since DM interaction prefers planar LRO and cuboc1 spin liquid since it energetically most favorable in absence of DM interaction. In all \textit{Ans\"{a}tze}, the magnitudes of $\mathcal{A}_{ij}$ and $\mathcal{B}_{ij}$ are same on all bonds, that is, 
$ \left | \mathcal{A}_{ij} \right | = \mathcal{A} $ and $ \left | \mathcal{B}_{ij} \right | = \mathcal{B} $. $\mathcal{A}$ and $\mathcal{B}$ are complex numbers and their phases  are  summarized in the caption of the Fig.~\ref{fig:ansatze} below. It is also assumed that the Lagrange's multiplier is independent of sites and is equal to $\lambda$. Using Fourier transformation, the Hamiltonian can be written as
\begin{equation}
H_{\text{MF}}=\sum_q \phi_q^\dagger N_q  \phi_q + N \big[  \lambda(2 S+1) + 2 ( \mathcal{A}^2 - \mathcal{B}^2)\big]
\end{equation}
where $N$ is the number of sites and  $\phi^\dagger_q = \big[ a_{1q}^\dagger , ...... a_{mq}^\dagger, b_{1,-q},......b_{m, -q} \big]$. The first index $\mu$ in $a_{\mu q}$ and $b_{\mu q}$ is the sublattice index.  $\mu$ can have values 1 to $m$ where $m$ is the number of sites per unit cell ($m=3$ for the $(0,0), (\pi,0)$ \textit{Ans\"{a}tze} and $m=6$ for \textit{cuboc1} \textit{Ansatz}) and $N_q$ is a $2m \times 2m $ matrix. 

This $N_q$ matrix can be diagonalised by  Bogoliubov  transformation $\phi_q = M_q  \xi_q $, where  $\xi^\dagger_q = \big[ \alpha_{1q}^\dagger , ...... \alpha_{mq}^\dagger, \beta_{1,-q},......\beta_{m,-q} \big]$. The transformation matrix must satisfy $M_q \tau M_q^\dagger=\tau $ where $\tau=\sigma_{3} \otimes  I_{m}$. We must find $M_q$ such that $M_q^\dagger N_q  M_q =\Omega_q$ where $\Omega_q = \text{diag} (\omega^\alpha_{\mu q}..........,\omega^\beta_{\mu,-q})$. The form of the $M_q$-matrix is given by $M_q= \begin{bmatrix}U_q & X_q \\ V_q & Y_q \end{bmatrix}$ where $U_q, V_q, X_q$ and $Y_q$ are the $m \times m$ matrix.

The diagonalised hamiltonian is given by
\begin{eqnarray}
H_{\text{MF}} & = &\sum_q \sum_\mu  \big( \omega_{\mu q}^\alpha \alpha_{\mu q}^\dagger \alpha_{\mu q} +\omega_{\mu q}^\beta  \beta_{\mu q} \beta_{\mu q}^\dagger  \big) \nonumber \\
& + & N \big[  \lambda(2 S+1) + 2 (\mathcal{A}^2 - \mathcal{B}^2)\big]
\end{eqnarray}
where $\omega^{\alpha}_{\mu q}$ and $\omega^{\beta}_{\mu q}$ are the excitation energies of $2m$ spinon modes. Note that for chiral \textit{Ansatz} $\omega_{\mu q}^\alpha \neq \omega_{\mu q}^\beta$. The ground state energy is given by
\begin{eqnarray}
E = \sum_{q,\mu}  \omega_{\mu q}^\beta +  N \big[  \lambda(2 S+1) + 2 (\mathcal{A}^2 - \mathcal{B}^2)\big]
\end{eqnarray}

Bogolioubov matrix is computed using the procedure outlined in ref~\cite{colpa1978diagonalization}. The mean field parameters $\mathcal{A}, \mathcal{B}, \lambda$, and $ \phi$ are found by extremizing the ground state energy $E$. The positive definiteness of the Hamiltonian puts several conditions on the domain  of these parameters. The method used for searching the saddle points is outlined in the next section. The complex link variables $\mathcal{A}_{ij}$ and $\mathcal{B}_{ij}$ satisfies the self consistency equations
\begin{equation}
\mathcal{A}_{ij}= \braket{ \hat{\mathcal{A}}_{ij} } \hspace{0.5cm} \text{\&}  \hspace{0.5cm} \mathcal{B}_{ij}=\langle \hat{\mathcal{B}}_{ij}\rangle
\end{equation}

which are equivalent to extremization of free energy as
\begin{eqnarray}
\frac{\partial E}{ \partial \mathcal{A}} =  \frac{\partial E}{ \partial \mathcal{B}} =  \frac{\partial E}{ \partial \mathcal{\phi}} = \frac{\partial E}{ \partial \lambda} = 0
\end{eqnarray}
\section{ Numerical search for saddle points}
The bosonic mean field Hamiltonian is diagonalizable if $N_q$ is positive definite for all $q$. We begin the search  by finding the valid domain in parameter space $\{\mathcal{A},\mathcal{B},\lambda\}$ where this conditions is true. The gapless LRO phases emerge as the 
corresponding saddle point approaches the boundaries, thus closing the gap in the spinon spectrum. These points are difficult to find since the eigenvalues of the Hessian has drastic varying magnitudes at the saddle point. As an example we have given the Hessian matrix of \textit{Ansatz} $(\pi, 0)$ at $S=0.366, \theta=0.21$ and $ N= 20$  
$$h=\left(
\begin{array}{ccc}
 0.9230 & -0.5794 & -1.0186 \\
 -0.5794 & -3.6344 & -0.3558 \\
 -1.0186 & -0.3558 & -9.9997\times 10^{11} \\
\end{array}
\right)$$
The eigenvalues of $h$ are given by $\{-9.9997\times 10^{11}, -3.7069, 0.9955 \}$. Thus in LRO phases, we rotate co-ordinate axes which are nearly parallel to the eigenvectors of the Hessian, two of which are parallel to the boundary surface. The main hurdle in this procedure is to finding boundary surfaces. However these can be guessed  by examining the classical orders. Thus analyzing eigenvalues of $N_q$ at few high symmetry points in reciprocal lattice is enough. For $\sqrt{3}\times \sqrt{3}$ phase,  the boundary plane is given by 
\begin{equation}
\sqrt{3}|\cos \theta| \mathcal{A} + \mathcal{B} + \lambda = 0
\end{equation}
\begin{figure}[ht!]
\includegraphics[height=4cm,width=4cm]{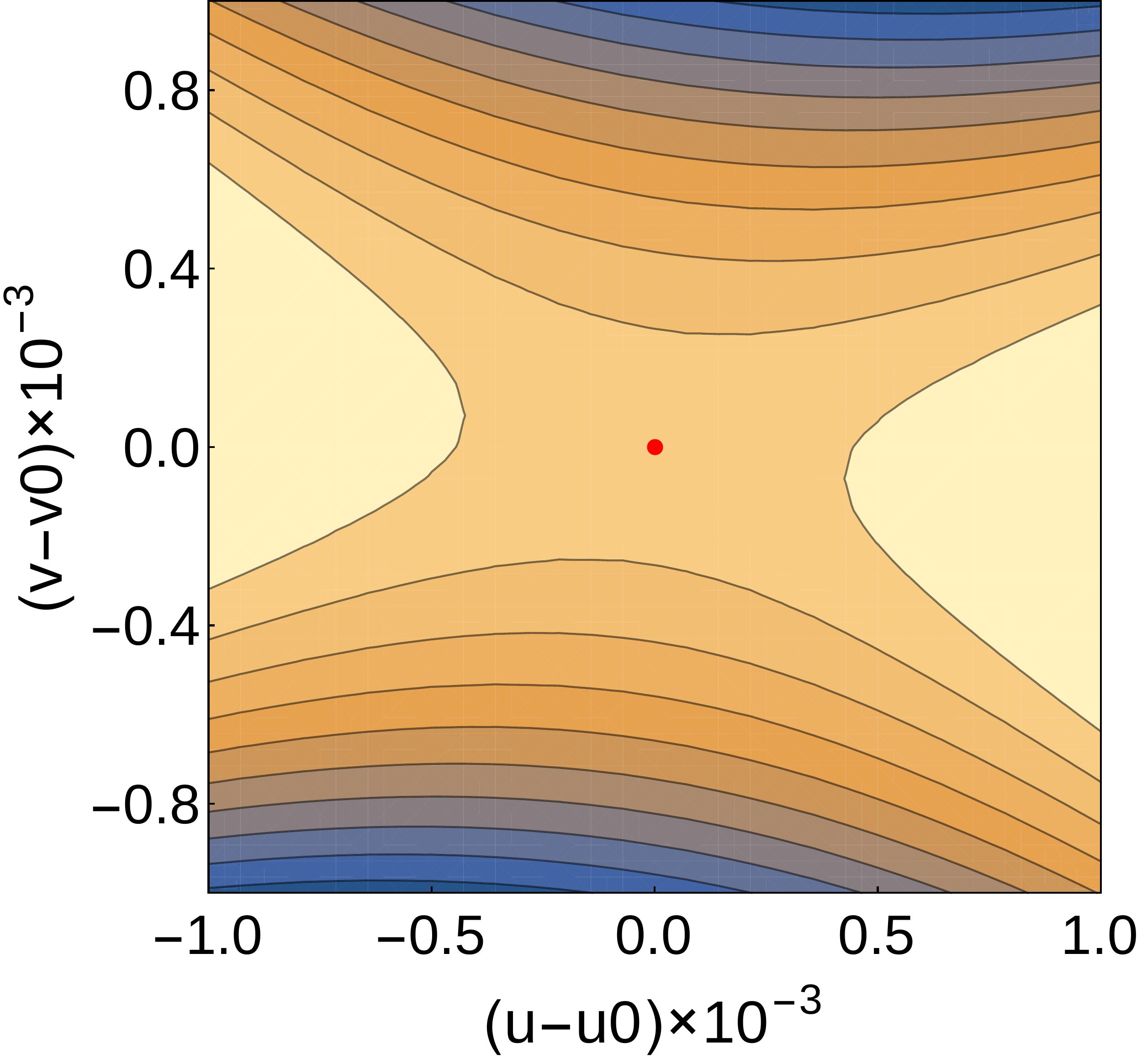}
\includegraphics[height=4cm,width=4cm]{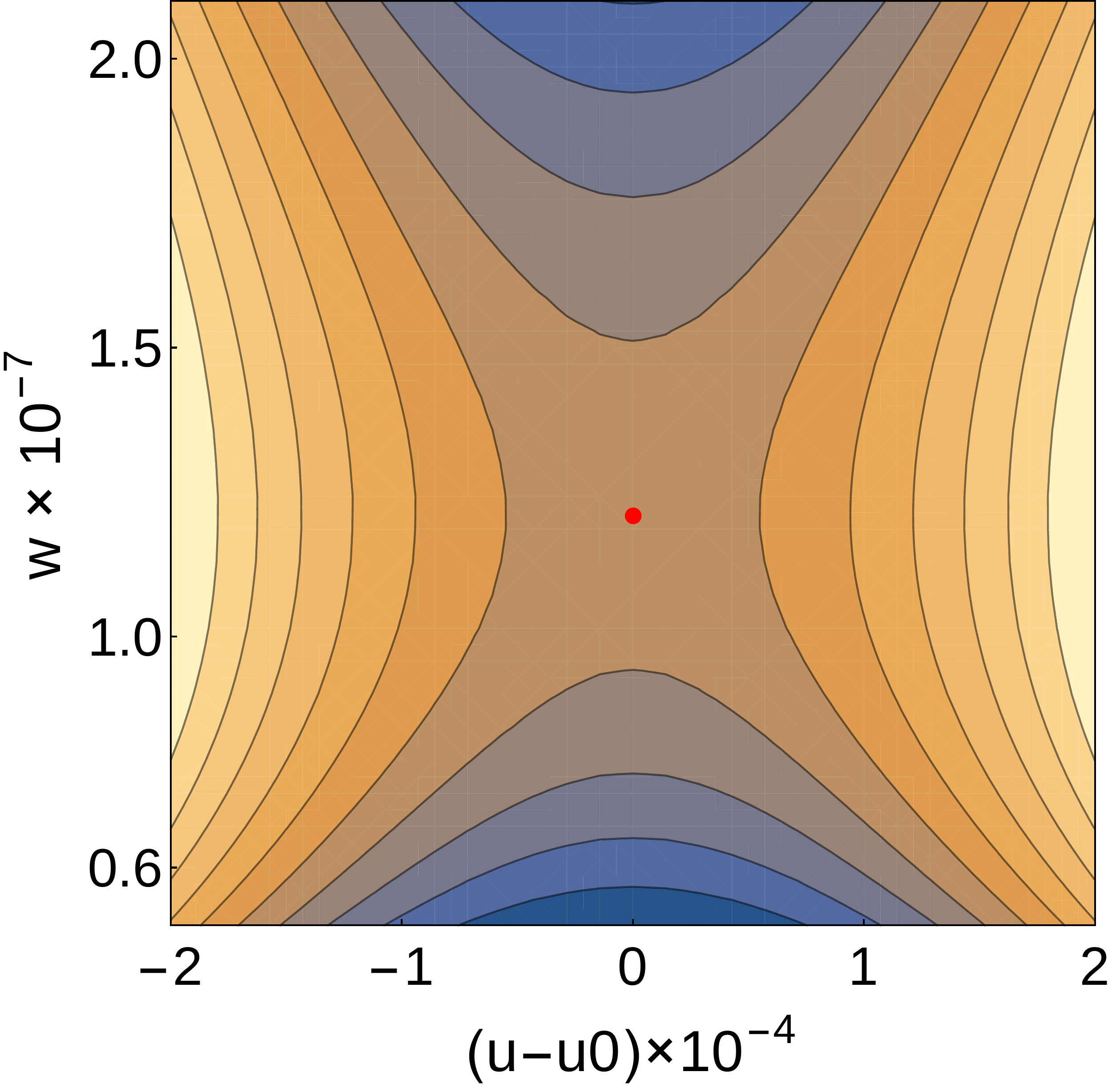}
\caption{Optimum points $(u0,v0,w0)$ in the rotated frame $(u, v, w)$ for \textit{Ansatz} $(\pi, 0)$ at $S=0.366$ and  $\theta=0.21$}
\end{figure}
The boundary surfaces in \textit{cuboc1} state are not planner but then it is still possible to search for saddle point along the surfaces and   then perpendicular direction. This way we have been able to achieve much better accuracy where the sum of squares of gradients is of the order of $10^{-14}$
\section{ PROPERTIES AND THE PHASE DIAGRAM}
\begin{figure}[ht!]
	\includegraphics[height=6cm,width=8cm]{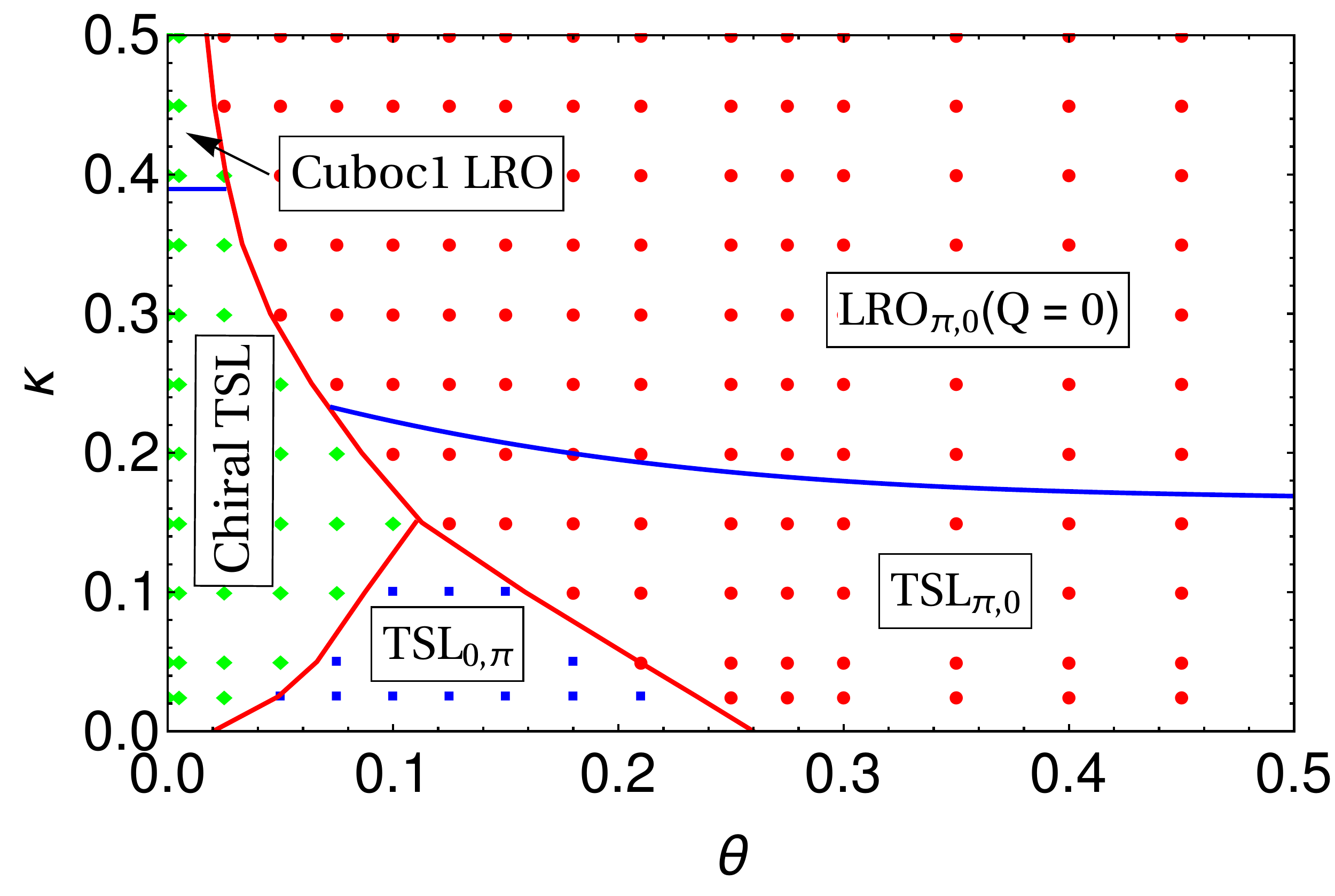}
	\caption{Ground State  phase diagram at $T=0$. In the phase diagram indicate second ordered phase transitions are shown in blue and the first ordered transitions by red lines.}
	\label{fig:phasediagram}
\end{figure}

\global\long\def\aa{\left(0,0\right)}
\global\long\def\ab{\left(0,\pi\right)}
\global\long\def\ac{\left(\pi,0\right)}
We have computed energies for various values of $\kappa\,(0\text{ to }0.5)$ and $\theta\,(0\text{ to }\pi /6)$ for all \textit{Ans\"{a}tze}. Based on the energy and the gap in the spinon spectrum, our proposed ground state phase diagram is shown in the Fig.~\ref{fig:phasediagram}. As expected, for low values of $\kappa$ and also small strengths of DM interaction, the phases are gapped spin liquids. For $\theta\lesssim0.5$, the gap in the spinon spectrum closes at about $\kappa=0.39$ indicating a second ordered transition from gapped liquid phase to gap-less cuboctahedral LRO. For large strength of DM interaction, that is for $\theta>0.2$, the system enters $\mathbf{Q}=0$ LRO phase at $\theta \gtrsim0.2$. Fig.~\ref{fig:Energy gap} shows that the gap closes at $\kappa=0.18$ for $\theta=0.3$. 

To illustrate the nature of various phases we also have calculated static spin structure factor 
\[
S^{\alpha\alpha}(\textbf{Q})=\frac{3}{4N}\sum_{i,j}q^{i\textbf{Q}\cdot(\textbf{R}_{i}-\textbf{R}_{j})}\langle0|S_{i}^{\alpha}S_{j}^{\alpha}|0\rangle
\]
 where $\alpha={x,y,z}$, $\textbf{R}_{i}$ and $\textbf{R}_{j}$ are the positions of the site $i$ and $j$. Due to DM
interaction, the global spin rotation symmetry is reduced to $U(1)$, and hence one must calculate both $S^{xx}$ and $S^{zz}$.
The expressions for the structure factor in terms of Bogoliubov matrices is given in appendix.

\subsection{Spin Liquid Phases}
\begin{figure}[ht!]
	\includegraphics[height=5.6cm,width=7.5cm]{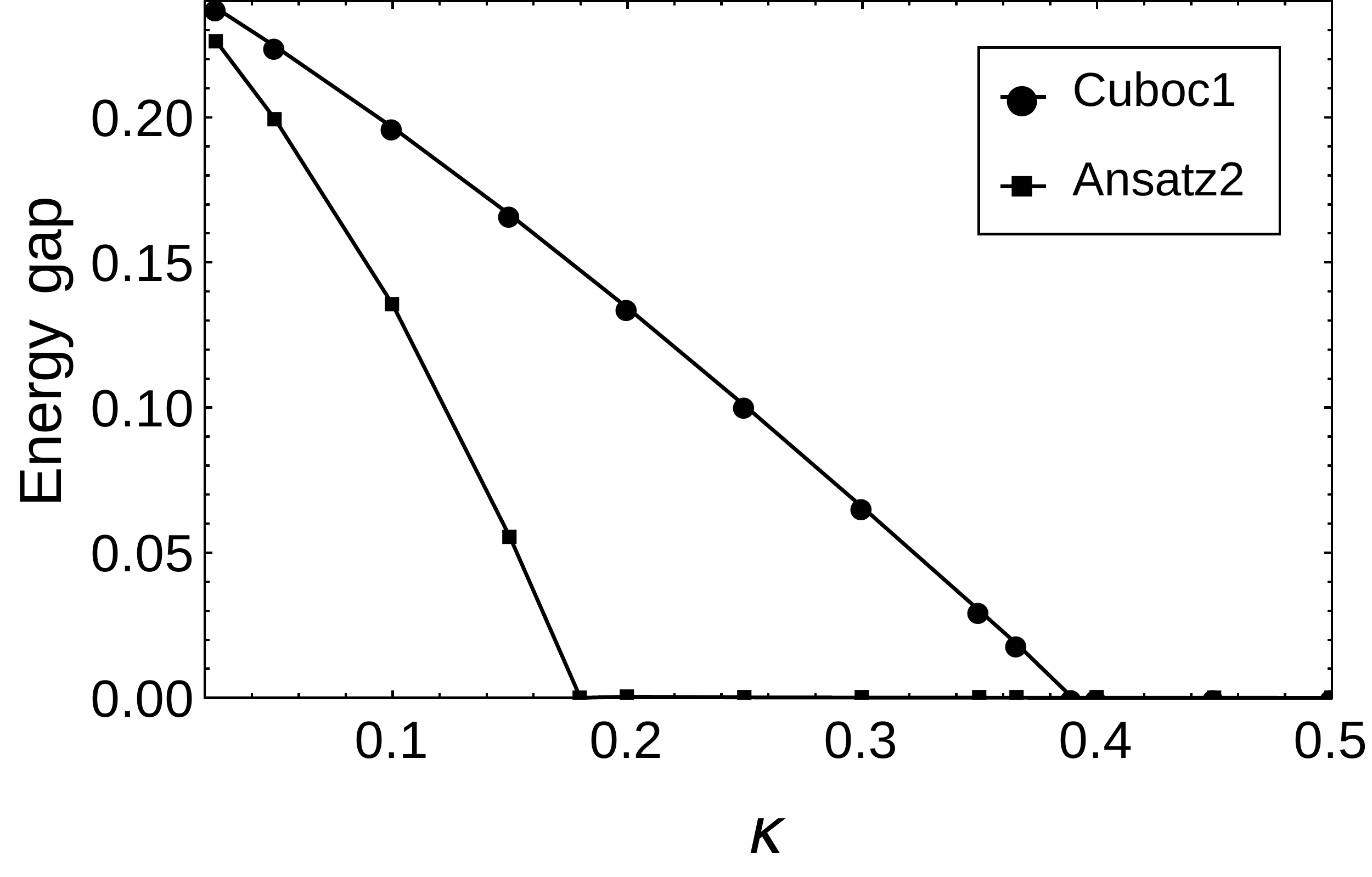}
	\caption{ Variation of energy gap with $\kappa$ for \textit{cuboc1} \textit{Ansatz} at $\theta = 0$ and  $(0,\pi)$ \textit{Ansatz} at $\theta = 0.30$}
	\label{fig:Energy gap}
\end{figure}

The phases of the isotropic Heisenberg model ($\theta=0$) have been studied by Messio and our results are matching with
them. For $\kappa>0.39$, the spinon spectrum is gap-less at $\mathbf{Q}=\frac{3}{2}\mathbf{K}$. At $\kappa\le0.39$, the
gap opens up and the system enters chiral spin liquid phase and remains in this phase down to $\kappa=0$. Fig.~\ref{fig:Energy gap} shows the variation of energy gap as a function of $\kappa$.  However it is interesting to note that the short range correlations are cuboctahedral near the phase transition but as $\kappa$ is decreased the peak in static structure factor becomes broad but also shift towards $\mathbf{K}$ point (Fig. 6(a)) before flattening out at $\kappa = 0.1$. Thus the short range correlations become more like $\sqrt{3}\times\sqrt{3}$ type as $\kappa$ is decreased which is consistent with $q_{1}=-q_{2}$ spin liquid state obtained
by Sachdev~\cite{huh2010quantum}.
\begin{figure*}[ht!]
	\includegraphics[height=4.8cm,width=6cm]{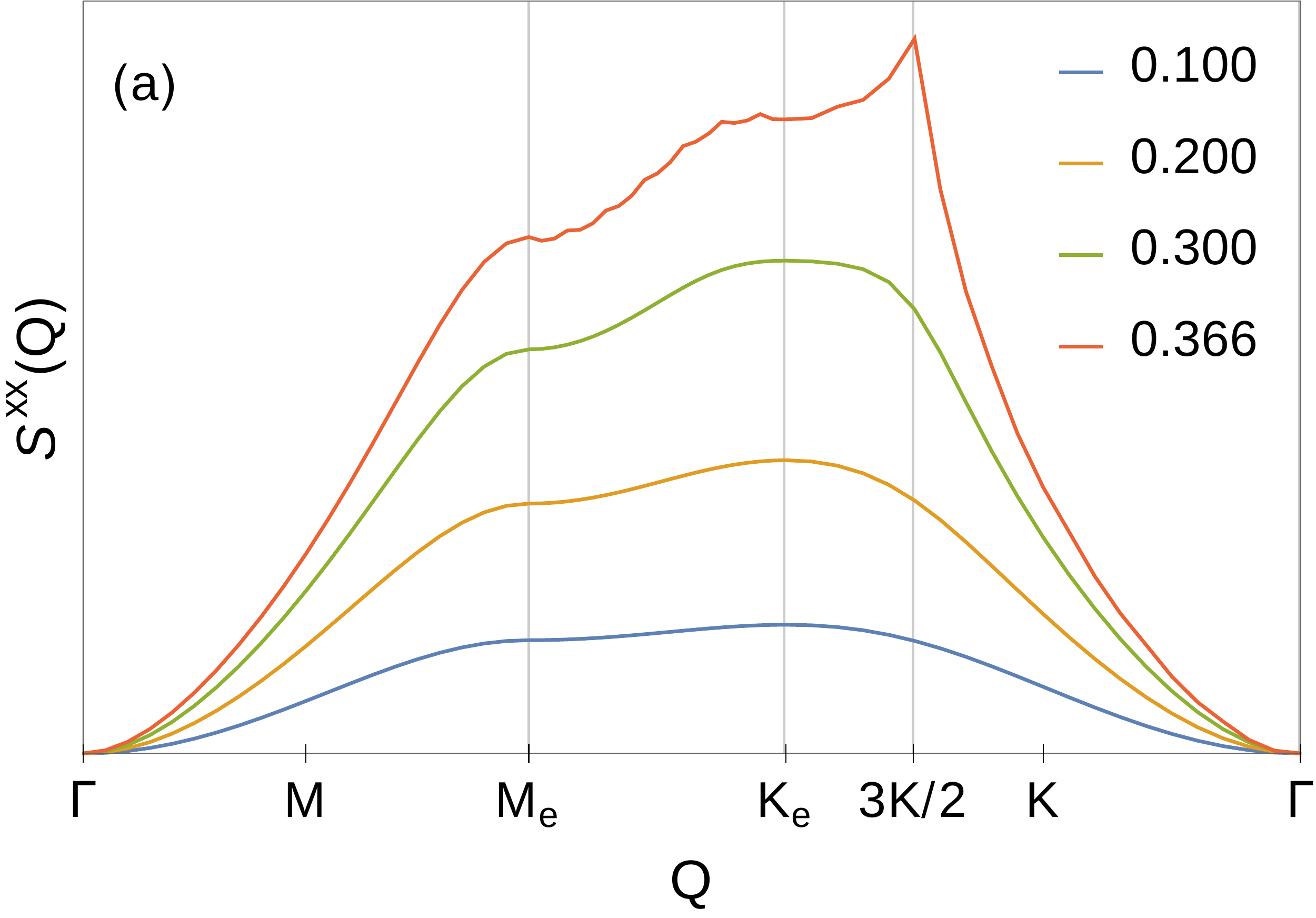}
	\hspace{0.1cm}
	\includegraphics[height=5cm,width=5.5cm]{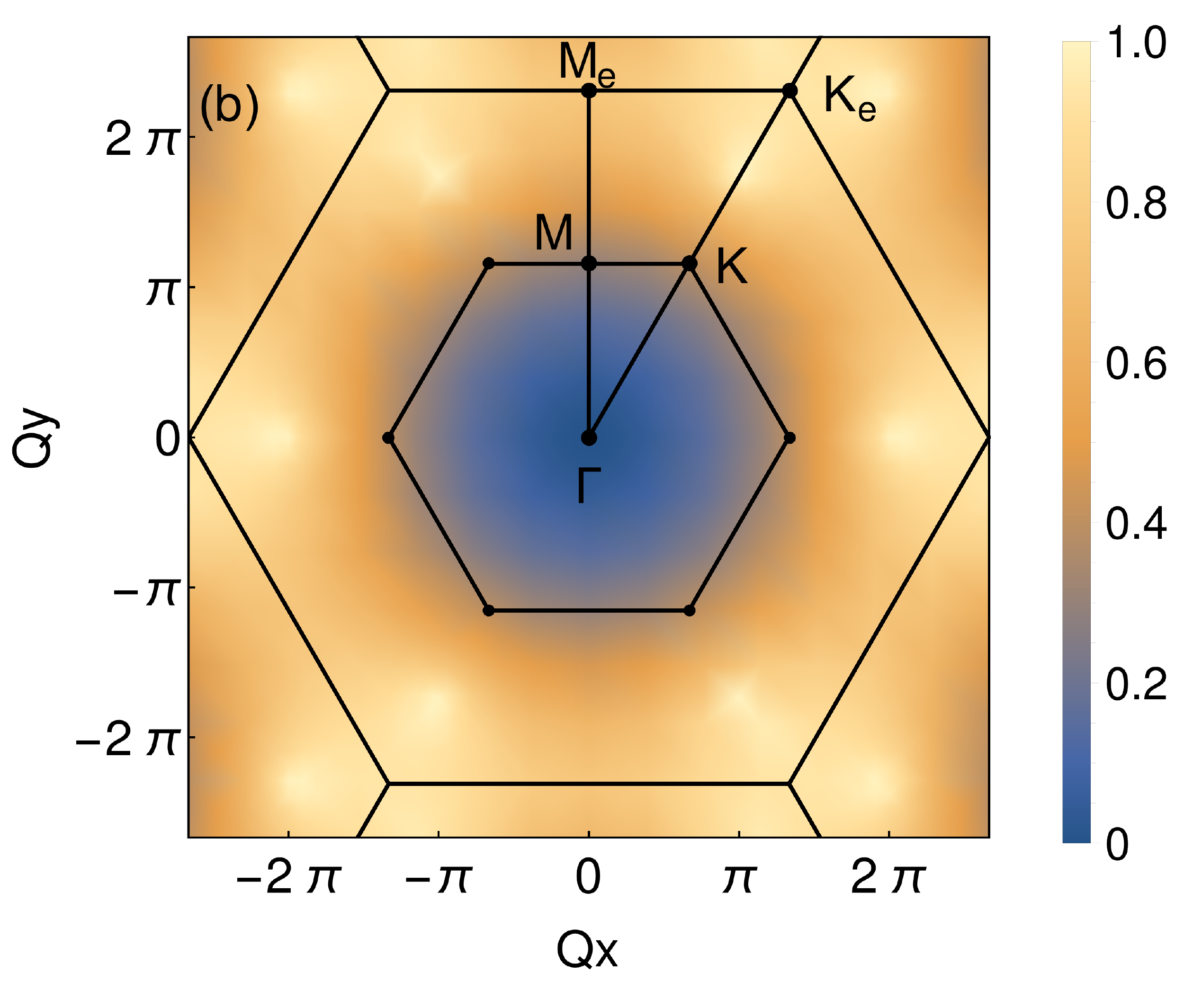}
	\hspace{0.1cm}
	\includegraphics[height=5cm,width=5.5cm]{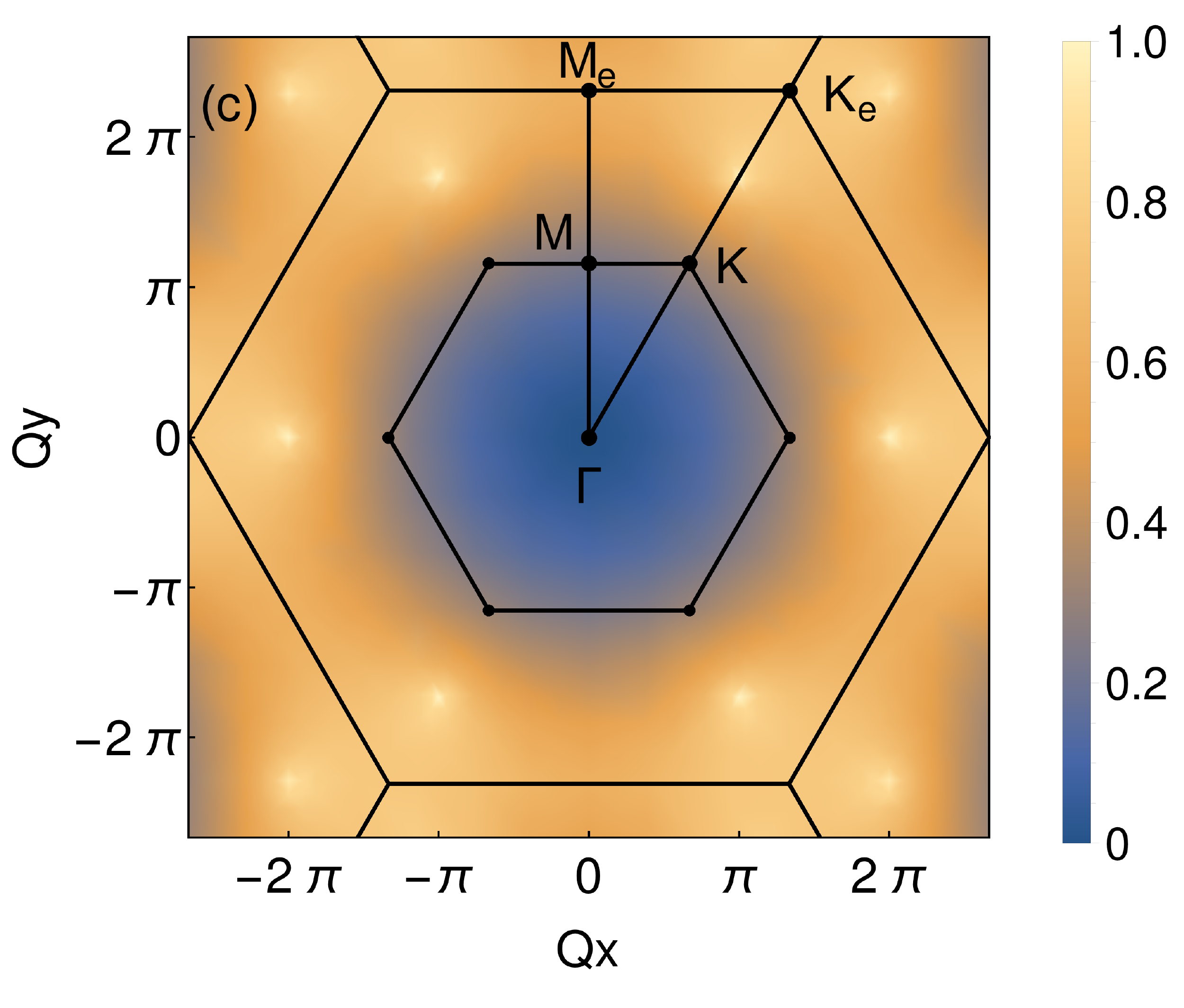}
	\caption{(a) XX-component of static  structure factor along the high symmetry line $\bf{\Gamma} - M_e - K_e -\Gamma$ for  \textit{cuboc1} \textit{Ansatz}  for different values of   $\kappa$ at $\theta=0$. XX-component of static  structure factor  for  \textit{cuboc1} \textit{Ansatz} at (b)  $S=0.366$ and $\theta=0$ (c) $S=0.5 $ and $\theta = 0$}
\end{figure*}

For small values of boson density, $\kappa(<0.15)$, the quantum fluctuations prevent any sort of long range order in the
system even in presence of strong DM interaction. For very small values of $\theta$, the system persists in \textit{cuboc1} spin
liquid phase before making a first order transition to $(0,\pi)$ spin liquid phase. In this phase, the static structure
factor again shows a very broad peak at $\mathbf{K}$ point. As the strength of the DM interaction is further increased the
system enters $(\pi,0)$ spin liquid phase with the short range correlations of type $\boldsymbol{Q}=0$ showing a broad
peak in spin structure factor at $\mathbf{M}_e$ point. Comparison with the phase diagram obtained by Messio, this transition
occurs at much smaller values of DM interaction. This indicates the inclusion of $\mathcal{B}$ fields helps stabilizing $(\pi,0)$
spin liquid against $(0,\pi)$ phase.
\subsection{Neel Ordered Phases}
For very small values of DM interaction and $\kappa>0.39$, all three \textit{Ans\"{a}tze} $\aa$, $\ab$ and \textit{cuboc1} are all in gapless  phases corresponding to classical orders $\sqrt{3}\times\sqrt{3}$ , $\boldsymbol{Q}=0$ and cuboctahedral respectively~\cite{messio2011lattice}. However the lowest energy in \textit{cuboc1} \textit{Ansatz}. In this phase, the soft modes in spinon spectrum are at $\frac{3}{4}\mathbf{K}$ points in extended Brillouin zone. Even though the \textit{Ansatz} is still symmetric, the mechanism of symmetry breaking leading to emergence long range order through bose condensation is well understood and is described in Sachdev~\cite{sachdev1992kagome} and Messio~\cite{messio2013time}. The spin structure factor in this phase shows strong peaks at $\frac{3}{2}\mathbf{K}$ point which confirms the \textit{cuboc1} LRO phase (Fig. 6(c)). At $\kappa=\frac{1}{2}$, the system is in \textit{cuboc1} LRO phase for $\theta < 0.0173$ and then enters $\boldsymbol{Q}=0$ phase with a first order phase transition. This width of \textit{cuboc1} LRO decreases as $\kappa\to\infty$ where even infinitesimal DM interaction immediately orders the system in planar configuration.

\begin{figure}[ht!]
	\includegraphics[height=3.6cm,width=4.25cm]{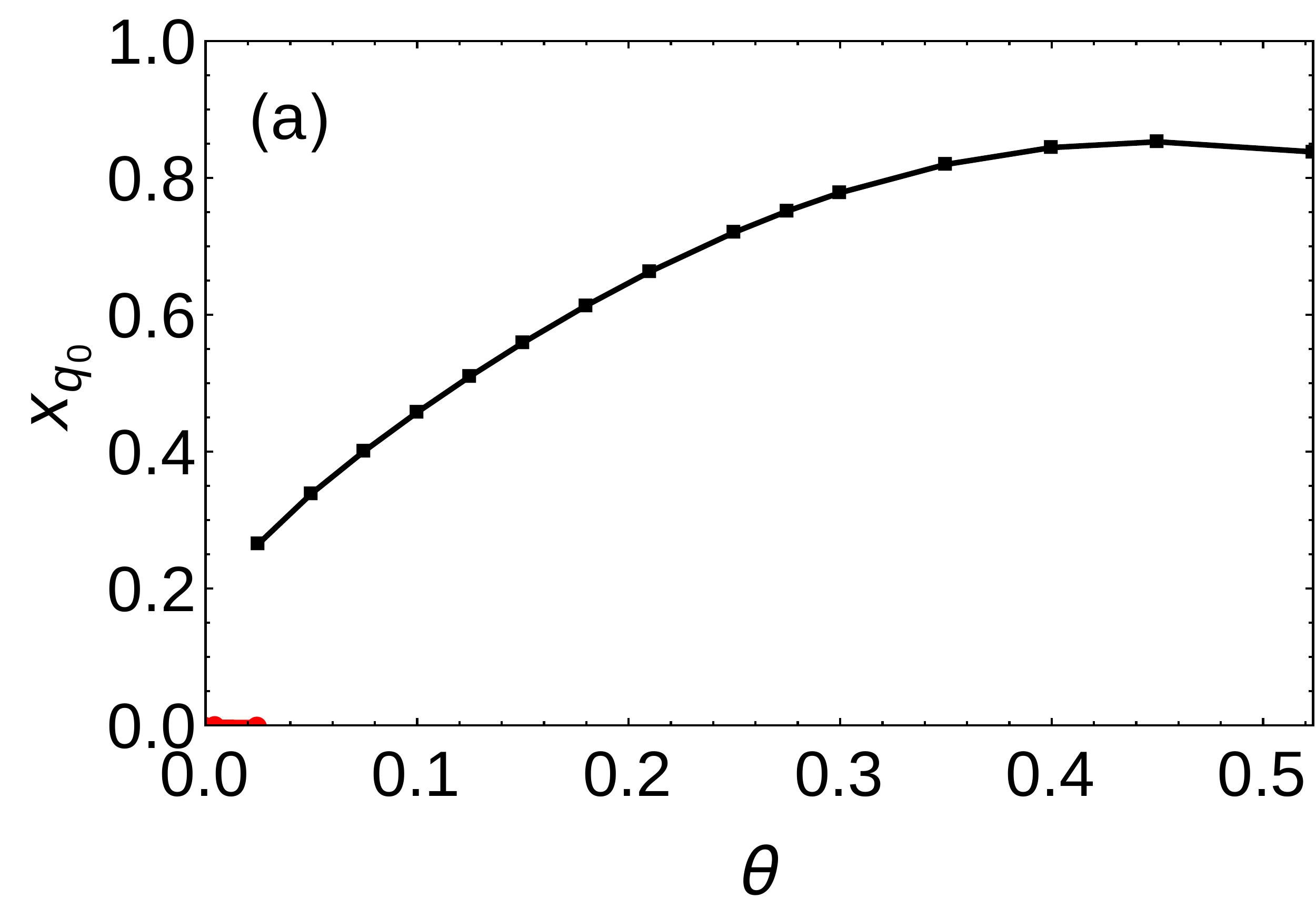}
	\includegraphics[height=3.6cm,width=4.3cm]{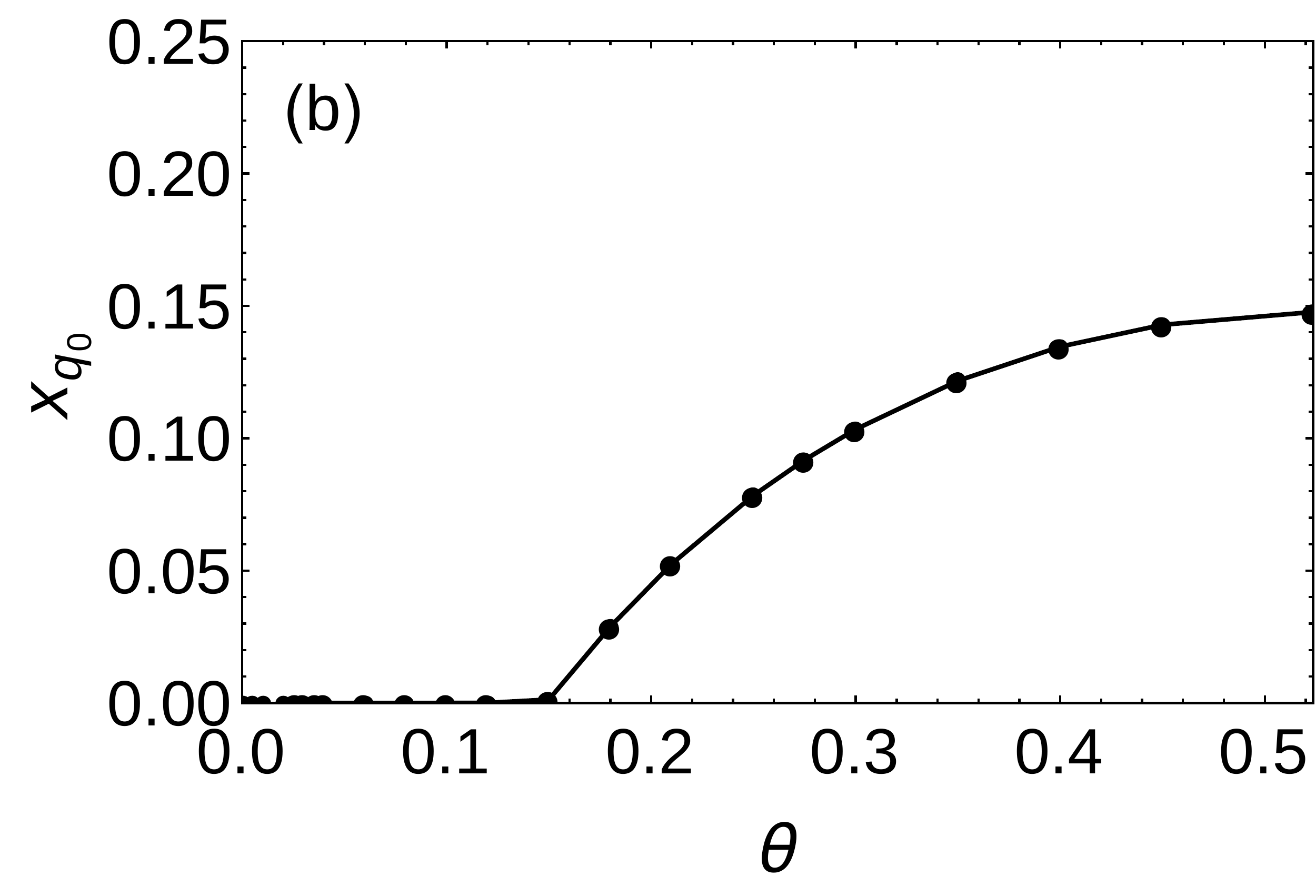}
	\caption{Condensate fraction $x_{q_{0}}$ as a function of $\theta$  (a) in the ground state at $\kappa=0.366$ where the red points shows \textit{cuboc1} \textit{Ansatz} and black points shows $(\pi,0)$ \textit{Ansatz} (b) for $\left(\pi,0\right)$ \textit{Ansatz} at $S=0.2$}
\end{figure}

For $\theta > 0.1$, the $\left(\pi,0\right)$ spin liquid shares a phase boundary with the $\mathbf{Q}=0$ phase, which is
favored by DM interaction. Across this boundary, the gap closes at $Q=0$. The signature sharp peaks in static structure
factor at $\mathbf{M}_e$ point. The condensate fraction $x_{0}$ in soft mode at $q_{0}$ can be computed using
\[
\frac{1}{2S}\sum_{i}\langle n_{i}\rangle=x_{q_{0}}N+\sum_{\substack{j,i \neq i_0{}\\q \neq q_0}}\frac{|V_{ij}(q)|^{2}}{S}.
\]
Fig. 7(a) shows the emergence of the condensate fraction as a function of $\theta$ along the $\kappa = 0.366$ horizontal line in the phase diagram. condensate fraction is zero in the \textit{cuboc1} \textit{Ansatz} and becomes nonzero for the $(\pi,0)$ \textit{Ansatz} via  first order phase transition at $\theta \approx 0.03$. In the Fig. 7(b), for $S=0.2$ the second order phase transition occurs near $\theta = 0.15$ for $(\pi,0)$ \textit{Ansatz}.
\subsection{Discussion}
The ground state phase diagram obtained here is qualitatively similar to that of Messio~\cite{messio2010schwinger} except that the \textit{cuboc1} \textit{Ansatz} replaces the $\aa$ \textit{Ansatz} in the phase diagram. The results obtained at $\kappa = 1/2$ however are not in agreement with earlier studies for spin 1/2 system~\cite{cepas2008quantum, laeuchli2009, huh2010quantum} which predict  a moment-free phase for small values of $\theta$ and  a second ordered phase transition to $\mathbf{Q}=0$ phase. However the nature of the short range correlations in liquid phase is unclear. Cepas et al argue that short range corrleations of type $\mathbf{Q}=0$, whereas Huh et al showed it to be of $\sqrt{3}\times\sqrt{3}$ type.  However, Messio~\cite{messio2012kagome} have argued that, in SBMFT, since $n_{i}=2S$ constraint is implemented only in an average sense, spin $1/2$ system may not be correctly represented by $\kappa=1/2$. Due to the fluctuations in $n_{i}$, the $\left\langle \boldsymbol{S}_{i}^{2}\right\rangle $ is overestimated to be $\frac{3}{2}S(S+1)$ at $\theta=0$. If we treat 
$\left\langle \boldsymbol{S}_{i}^{2}\right\rangle $ to be a good quantum number, then the spin $1/2$ system is approximated by the bosonic  system at $\kappa=0.366$. At this value of $\kappa$, our proposal shows that the system is $\mathbb{Z}_2$ chiral spin liquid till 
 $\theta < 0.03$ with short range correlations of cuboctahedral kind. For $\theta > 0.03$, the DM interaction forces the spins to be in planar arrangement with $\boldsymbol{Q}=0$ long range order.
 
 It is also possible that the best representation may be at smaller values of $\kappa$, where we have noted that the short range correlations  tend to be more like $\sqrt{3}\times\sqrt{3}$ type as argued by Messio et al.  Even though at $\kappa = 0.366$ shows a first order transition from spin liquid to $\mathbf{Q} = 0$ LRO, it may not be adequate in reconciling with the experimental results. The strength  of the DM interaction is estimated to be $0.08J$ in $\mathrm{ZnCu}_3(\mathrm{OH})_6\mathrm{Cl}_3$. This compound does not exhibit  freezing of magnetic moments to very low temperatures~\cite{helton2007spin}.  In the present mean field theory, at $\kappa = 0.366$, the critical $D_c = 0.03J$.  The optimized value of mean field parameters and energy for $S=0.366$ is as given in the table below.
\begin{table}[ht]
	\centering 
	\begin{tabular}{|c| c| c|c| c| c| c|} 
		\hline
 		\textit{Ansatz} & $\theta$ & $\mathcal{A}$ & $\mathcal{B}$ &$\lambda$ &$\phi$ & energy\\ [0.5ex] 
		\hline 
		cuboc1& 0 &0.4036& 0.1185&-0.5803&1.9847&-0.2976 \\
		\hline
		{ $(\pi, 0)$} & 0.21 & 0.4226 & 0.1340&-0.6700& - & -0.3213 \\ [1ex] 
		\hline
	\end{tabular}
	\caption{Optimized values and energies for different \textit{Ans\"{a}tze}}
	\label{table:nonlin} 
\end{table}
\section{Dynamical spin  structure factor}
\begin{figure*}[ht!]
\includegraphics[height=4.7cm,width=5.5cm]{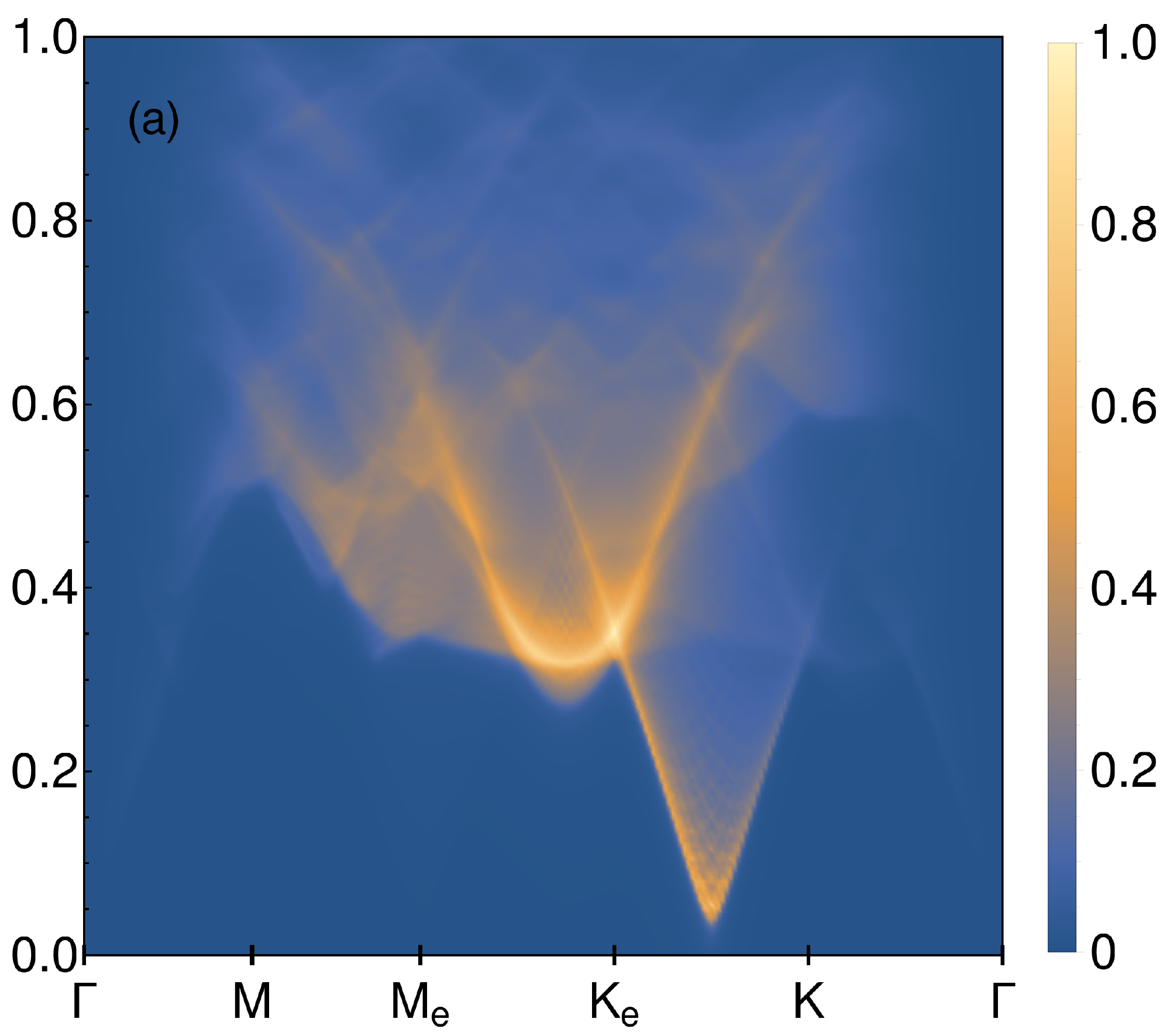}
\hspace{0.5cm}
\includegraphics[height=4.7cm,width=5.5cm]{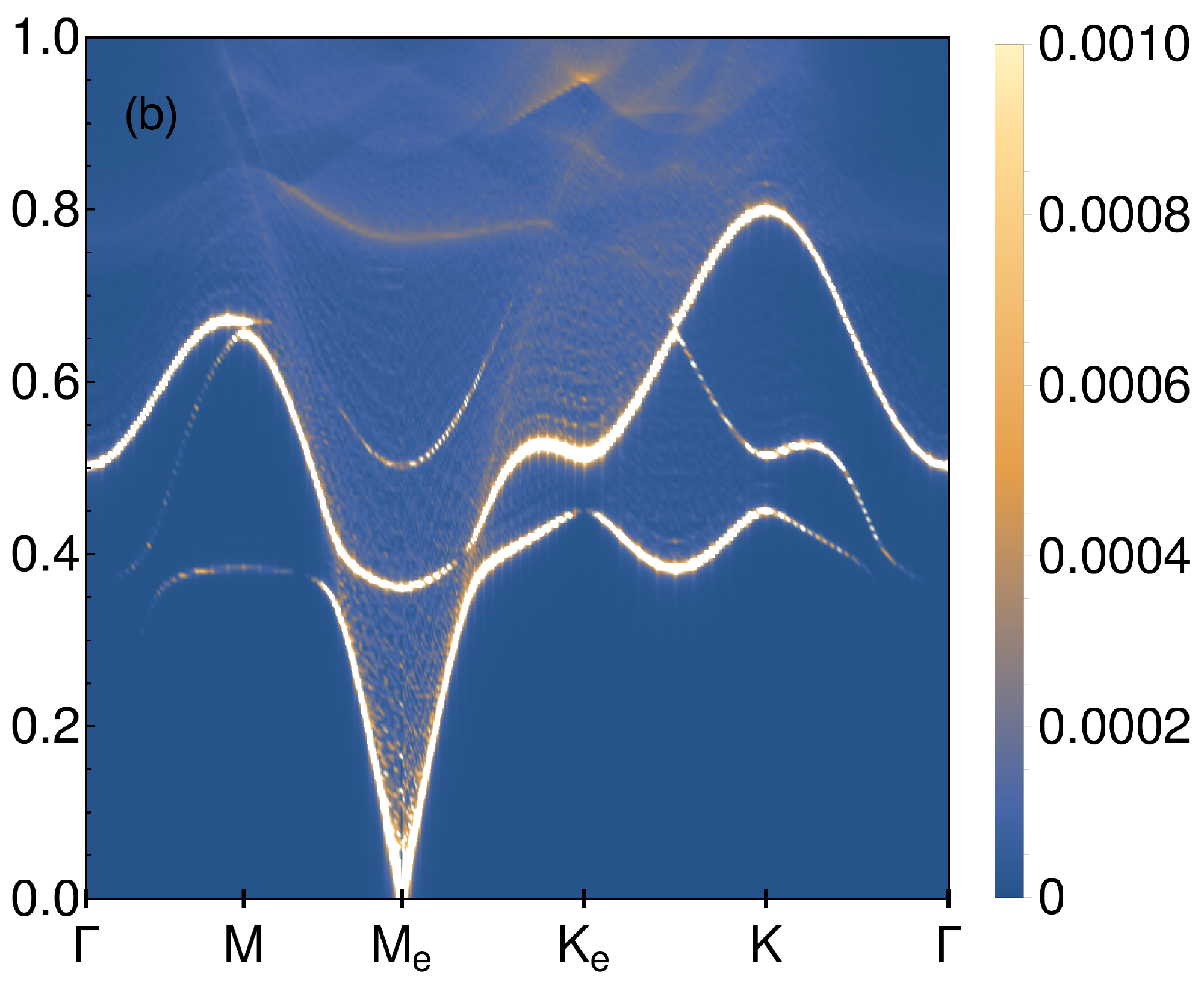}
\hspace{0.5cm}
\includegraphics[height=4.7cm,width=5.5cm]{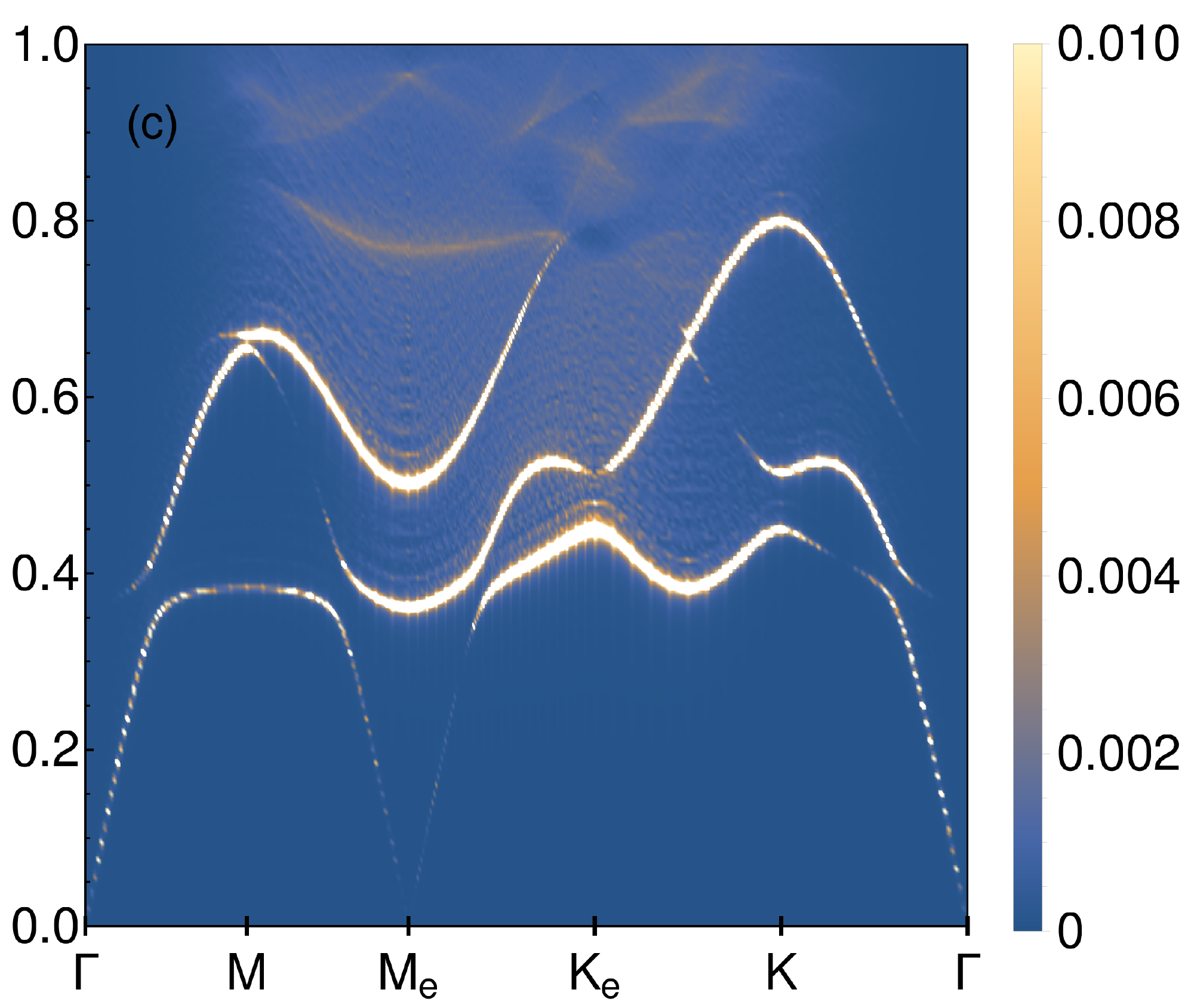}
\caption{XX-component of dynamical structure factor for (a) $cuboc1$ \textit{Ansatz} at $S=0.366$ and $\theta = 0 $ (b) $(\pi,0)$ \textit{Ansatz} at $S=0.366$ with $\theta = 0.21 $ (c) ZZ-component of dynamical structure factor for $(\pi,0)$ \textit{Ansatz} at $S=0.366$ and $\theta=0.21$. In LRO phase we choose scale to be very small such that the spinon continuum is visible. }
\end{figure*}
Since in neutron scattering experiment the inelastic  neutron scattering cross section is proportional to the dynamical spin structure factor, we have calculated the dynamical spin structure factor defined by 
\begin{equation}
S^{\alpha \alpha} (\textbf{Q},\omega) =\int_{ - \infty}^{  \infty} \langle 0| S_{-\textbf{Q}}^\alpha (t) S_{\textbf{Q}}^\alpha (0)| 0  \rangle e^{i \omega t} dt  
\end{equation}
where $\alpha = {x, y, z}$ and  $|0 \rangle$ is the ground state of the system . The expressions for ZZ  and XX component of dynamical structure factor in terms of Bogoliubov matrices are given in the appendix. For numerical purpose, the delta function is approximated by Lorentzian function. The results are qualitatively similar to the results obtained by Messio.

In the figure, we illustrate the evolution of the  dynamical structure factor across the phase transition from spin liquid to LRO phase for $\kappa = 0.366$.  The dynamical structure factor is shown along the high symmetry line $\mathbf{\Gamma} - \mathbf{M}_e - \mathbf{K}_e - \mathbf{\Gamma}$. The dynamical structure factor of \textit{cuboc1} \textit{Ansatz} at $\theta = 0 $ in Fig. 8(a) shows sharp onset of spinon continum at $\frac{3}{2} \mathbf{K}$ with very small gap. This is due to the proximity of critical point at $\kappa=0.39$ beyond which there is \textit{cuboc1} long range order.  

Since the DM interaction reduces the global spin rotation symmetry from $SU(2)$ to $U(1)$, there is notable difference between XX and ZZ component of dynamical structure factor.  In Fig. 8(b) and Fig. 8(c), we show  the dynamical structure factor of \textit{Ansatz} $(0,\pi)$  at $\theta = 0.21$ which is $\mathbf{Q} = 0 $ LRO phase.  

In the XX-component of dynamical structure factor,  the magnon branches are quite strong and   in the the density plot, it is visible as  three lines of bright dots corresponding to three magnon branches.  These magnon branches are obtained by considering the pair of spinons, one in the soft modes and other one from the excited state. The elastic peak (along $\omega = 0 $ line) appears with largest intensity at $ \mathbf{M}_e $ point which is the $\mathbf{\Gamma}$ point of the next wigner- seitz cell. 

 In the ZZ-component of dynamical structure factor the magnon branches are of very low intensity and  suppressed, as the spins are forced to lie in the XY plane due to the DM interaction.  

\section{Conclusion }	
We have computed the ground state phase diagram of Heisenberg Kagome antiferromagnet with DM interaction using SBMFT approach. We have included the time reversal symmetry breaking chiral \textit{Ansatz} proposed by Messio~\cite{messio2012kagome}and also considered hopping mean field $\mathcal{B}$.  For large $S$, even with small DM interaction, the spins are forced to in plane and in $\mathbf{Q} = 0$ long range order. For small $S$, the quantum fluctuations induce series of spin liquid phases with increasing DM interaction. In this region, the inclusion of the hopping field seems to stabilize $(\pi,0)$ spin liquid phase over $(0, \pi)$ spin liquid.

Since the constraint of boson density is implemented strictly, $\braket{\mathbf{S}_i^2} = \frac{1}{2}$ at $\kappa = 0.366$ due to the fluctuations in boson density. We find that, at $\kappa = 0.366$, the model shows a first ordered phase transition from chiral \textit{cuboc1} spin liquid to $\mathbf{Q} = 0$ Neel phase at $D=0.03 J$.  Even though, this result is qualitatively in agreement with other studies, it is not adequate in explaining the moment free phase of Herbertsmithite~\cite{helton2007spin} since the estimated DM strength is $0.08 J$. Probably, one may need to consider even smaller values of $\kappa$ to obtain better numerical agreement with other studies and experiment. We, have also calculated static and dynamical structure factor at the representative $\kappa = 0.366$ point. 
\section{APPENDIX }
\subsection{Static spin structure factor}
The expression for XX-component of static spin structure factor is given by
\begin{eqnarray}
S^{xx}(\mathbf{Q}) & = & \frac{3}{16N} \sum_{q , \mu \nu} \Big[ [X_{\mathbf{Q} + q}^* Y_{\mathbf{Q} + q}^T]_{\mu \nu} [ Y_{-q}^* X_{-q}^T]_{\mu \nu} \nonumber \\
 & + & [X_{\mathbf{Q} + q}^* X_{\mathbf{Q} + q}^T]_{\mu \nu}  [ Y_{-q}^* Y_{-q}^T]_{\mu \nu}\nonumber \\
& + & [V_q U_q^\dagger]_{\mu \nu} [U_{-\mathbf{Q}-q} V_{-\mathbf{Q}-q}^\dagger]_{\mu \nu}\nonumber \\
&+ & [V_{-\mathbf{Q}-q} V_{-\mathbf{Q}-q}^\dagger]_{\mu \nu} [U_q U_q^\dagger]_{\mu \nu}
 \Big] 
\end{eqnarray}
 The ZZ- component of structure factor has the form
\begin{eqnarray}
S^{zz}(\textbf{Q}) & =& \frac{3}{16N}\sum_{q,\mu, \nu} \Big[ [X_{{\textbf Q}+q}^* X_{{\textbf Q}+q}^T]_{\mu \nu} [U_q U_q^\dagger]_{\mu \nu} \Big] \nonumber \\
&+& [V_{-{\textbf Q}-q} V_{-{\textbf Q}-q}^\dagger]_{\mu \nu} [X_{-q}^*X_{-q}^T ]_{\mu \nu}\nonumber \\
&-&  [X_{{\textbf Q}+q}^* Y_{{\textbf Q}+q}^T]_{\mu \nu}[U_q V_q^\dagger]_{\mu \nu}\nonumber \\
&-& [V_{-{\textbf Q}-q} U_{-{\textbf Q}-q}^\dagger]_{\mu \nu}  [Y_{-q}^*X_{-q}^T ]_{\mu \nu}
\end{eqnarray}
\subsection{Dynamical spin structure factor}
The expression for XX-component of dynamical spin structure factor is given by
\begin{eqnarray}
S^{xx} (\textbf{Q},\omega) &=& 2 \pi \sum_q \sum_{ \mu \nu}\Big[|[U_q^\dagger V_{-q-{\textbf Q}}^*]_{\mu \nu} + [U_{-q-{\textbf Q}}^\dagger V^*_{q}]_{ \nu \mu}|^2 \nonumber \\
& + &|[Y_{-q}^T X_{q+{\textbf Q}}]_{\mu \nu} +[ Y_{{\textbf Q}+q}^T X_{-q}]_{\nu \mu}|^2 \Big]\nonumber \\
&\times& \delta(\omega - \omega_p)
\end{eqnarray}
where $\mu, \nu$ are the sublattice index. The expression for ZZ-component of dynamical spin structure factor reduces to
\begin{eqnarray}
S^{zz} (\textbf{Q},\omega)& = & 2 \pi \sum_q \sum_{ \mu \nu}| \big[ U_q^\dagger X_{{\textbf Q}+q} \big]_{\mu \nu} - \big[V_q^\dagger Y_{{\textbf Q}+q}\big]_{\mu \nu}|^2 \nonumber \\
&\times&  \delta(\omega - \omega_p)
\end{eqnarray}
where $\omega_p=\omega_q + \omega_{-{\textbf Q}-q}$. At ${\textbf Q}=0$ the $zz$-component will straight way will give zero because it is just the dot product of the two columns of the M matrix(para orthogonalization), consistent with the figure.
%

\end{document}